\documentclass[12pt]{article}
\usepackage{amssymb}
\catcode`@=11

\ifcase \@ptsize
    q mods for 10 pt
\or 
\or 
\fi

\newtheorem{theorem}{Theorem}[section]
\newtheorem{lemma}[theorem]{Lemma}

\newtheorem{proposition}[theorem]{Proposition}

\newtheorem{definition}[theorem]{Definition}

\newtheorem{example}{Example}
\newtheorem{fact}{Fact}

\newcommand{\beginproof}{\medskip\noindent{\bf Proof.~}}
\newcommand{\beginproofof}[1]{\medskip\noindent{\bf Proof of #1.~}}
\newcommand{\finishproof}{\hspace{0.2ex}\rule{1ex}{1ex}}
\newenvironment{proof}{\beginproof}{\unskip\nolinebreak\finishproof\par\medskip }

\title{Entropy lower bounds of quantum decision tree complexity
      \footnote{This work was supported in part by the National Science Foundation 
under grant CCR-9820855.}}
\author{Yaoyun Shi\footnote{ Computer Science Department,
			Princeton University,
			Princeton, New Jersey 08544.
			E-mail: shiyy@cs.Princeton.EDU.}}
\date{August 22, 2000}
\begin{document} 
\maketitle
\begin{abstract}\noindent
We prove a general lower bound of quantum decision tree
complexity in terms of some entropy notion.
We regard decision tree computation as a communication
process in which 
the oracle and the computer
exchange several rounds of messages, each round consisting of 
$O(\log{n})$ bits. 
Let $E(f)$ be the Shannon entropy of the random
variable $f(X)$, where $X$ is taken
uniformly random in $f$'s domain. Our main result is
that it takes $\Omega(E(f))$ queries 
to compute any \emph{total} function $f$. 
It is interesting to contrast this bound with the 
$\Omega(E(f)/\log{n})$ bound, which is tight
 for \emph{partial} functions. Our approach is
the polynomial method.
\end{abstract}

\emph{keywords:} Quantum computation; Decision tree; 
Lower bounds; Computational complexity; Entropy

\section{Introduction}
The decision tree model is probably the simplest 
model in the study of computational complexity. 
In this model,
the input $x$ is known only to an oracle,
and the only way that the computer can
 access the input is to ask
 the oracle questions of the type
`$x_i = ?$'. The computational cost is simply
the number of such queries, and the complexity of a 
problem is the minimal worst case cost. 
For example, to find out
whether or not there is a $1$ in $x_1, x_2, \cdots, x_n$,
any deterministic decision tree algorithm needs to
ask for all the $x_i$'s in the worst case. Therefore, its
(deterministic) decision tree complexity is $n$. 

Unlike \emph{classical} decision trees, 
a \emph{quantum} decision tree algorithm 
can make queries in a quantum \emph{superposition}, 
and therefore may be intrinsically faster
than any classical algorithm. For example, Grover's 
quantum algorithm\cite{Grover96}
for finding the location of the only $1$ in an $n$ bit string
makes only $O(\sqrt{n})$ queries, while any classical
algorithm needs $\Omega(n)$ queries.
In recent years, the quantum 
decision tree model has been extensively studied 
by many authors from both upper bounds and lower bounds
perspectives. Here we consider the latter aspect only.
\cite{BuhrmanW99} is a recent survey of
both classical and quantum decision tree complexity.

Throughout this paper, 
$f$ denotes a function: 
$$f: \{0, 1\}^n \supseteq A 
\rightarrow B = f(A) \subseteq \{0, 1\}^m,$$ 
for some integers $n, m > 0$.
Let $Q(f)$ be the quantum decision tree complexity
of $f$ with error probability bounded by 
$1/3$. Our goal is to derive a general lower bound 
for $Q(f)$ in terms of $E(f)$ defined as follows:

\begin{definition}
For any $f$, define the \emph{entropy}
of $f$, $E(f)$, to be the Shannon entropy of $f(X)$, where 
$X$ is taken uniformly random from $A$. More explicitly,
$$E(f) = \sum_{y \in B} p_y \log_2 \frac{1}{p_y},$$ 
where $p_y = \mathrm{Pr}_{x \in_R A}[ f(x) = y ]$.
\end{definition}

We first note the following general lower bound:
\begin{proposition} \label{logfactor}
For any $f$, $Q(f) = \Omega(E(f)/\log{n})$.
\end{proposition}

This fact can be proved by a standard information theoretical
argument, which we sketch here.
The computation can be viewed as
a process of communication: to make a query,
the algorithm sends the oracle $\lceil\log_2{n}\rceil +1$ bits, 
which are then returned by the oracle. 
The first
$\lceil\log_2{n}\rceil$ bits specify the location of the
input bit being queried and
the remaining one bit allows the oracle to write down the answer.
Now we run the algorithm on 
$\frac{1}{\sqrt{|A|}} \sum_{x\in{}A} 
|x\rangle_X |\overrightarrow{0}\rangle_Y$, where
$X$ and $Y$ denote the qubits that hold the input and the intermediate
results of the computer respectively. Now we consider $S_B^{(t)}$,
the \emph{von Neumann entropy} of qubits in $Y$ after the $t$th
query. If the algorithm computes $f$
in $T$ queries, at the end of the computation,
we expect to have a vector {\em close} to
$\frac{1}{\sqrt{|A|}} \sum_{x\in{}A}|x\rangle_X |f(x)\rangle_Y.$
Clearly $S_B^{(0)}=0$, $S_B^{(T)} \approx E(f)$,
and $|S_B^{(t+1)}-S_B^{(t)}| = O(\log{n})$ for 
any $t$, $0\le t\le T-1$.
The latter two assertions can be proved by  standard applications of
Holevo's theorem\cite{Holevo73}.
Therefore $T = \Omega(E(f)/\log{n})$. We will
 provide an example later 
to show that indeed this bound is tight. This means
one quantum query can get $\log{n}$ bits of information,
while any classical query can only get no more than
$1$ bit of information.

Surprisingly, this power of getting $\omega(1)$
bits of information in a query is not useful
in computing
{\em total} functions, i.e., functions that
are defined on every string in $\{0, 1\}^n$, in the
sense that each quantum query can only
get $O(1)$ bits of information {\em on average}, 
as stated in our main theorem:

\begin{theorem}[Main Theorem]
For any total function $f$, $Q(f) = \Omega(E(f))$.
\end{theorem}

Now we sketch the proof idea.
We take the polynomial approach initiated 
in \cite{BealsBCMW98}. Any correct algorithm that computes
$f$ will produce a set of polynomials 
$$\{\tilde{f}_y : \{0, 1\}^n \rightarrow \mathbb{R}: y\in B\}.$$
Each $\tilde{f}_y$ is an approximation to 
the characteristic polynomial for
$f^{-1}(y)$. If $f$ is a total function, on any Boolean
inputs, $\tilde{f_y}$ is forced to be close to either $0$ or
$1$, and this `take-it-or-leave-it' nature makes
it harder to approximate; in contrast, when 
$f$ is not a total function, on inputs where $f$ is not
defined, $\tilde{f}_y$ has more freedom to
take values that make the approximation
easier.

There are several previous papers that prove general 
lower bounds on quantum decision tree complexity in terms
of different complexity notions: \cite{BealsBCMW98}
by Boolean (block) sensitivity and by 
degree of approximating polynomials, \cite{Ambainis00}
by a combinatorial property, and \cite{Shi00} by
average Boolean sensitivity. 

In the next two sections we shall provide a rigorous definition of
the quantum decision tree model and then prove the main theorem.

\section{Quantum decision tree model}
In the quantum decision tree model, the computer 
has three sets of qubits: $P$, $Q$, and $R$. $P$
has $n$ bits, which hold the input; $Q$
has $\lceil\log_2{n}\rceil + 1$ bits, which 
contain a pointer to the input bits (i.e.,
an integer between $1$ and $n$),
as well as one more bit; R has an unlimited number of 
bits which serve as the algorithm's working space. 
A quantum decision tree computation with input $x$ is 
the application (from the right to the left)
 of a sequence of unitary operators
$$A := U_T O U_{T-1} O \cdots U_1 O U_0$$
on the initial state 
$$|x\rangle_P|\overrightarrow{0}\rangle_{QR},$$
where $O$ is the \emph{oracle gate}:
$$O |x\rangle_P |i, b\rangle_Q |c\rangle_R= |x\rangle_P |i, 
b\oplus\\x_i\rangle_Q |c\rangle_R,$$
and each $U_t = I\otimes\tilde{U}_t$, $0\le t \le T$, 
where $I$ is the identity operator on $l_2(P)$
and $\tilde{U}_t$ a unitary operator
on $l_2(Q\cup R)$.
We say that the algorithm computes $f$ (with error
bounded by $1/3$) if there exists a measurement
M on $l_2(Q\cup{}R)$, such that
for any $x\in A$, with probability no less than $2/3$
$f(x)$ will be observed by applying $M$
at the final state of the computation.
The quantum decision tree complexity $Q(f)$ 
is defined to be the minimal $T$ such that
there is a quantum decision tree algorithm 
that computes $f$ in $T$ queries.

The following example
demonstrates that the lower bound in
Proposition~\ref{logfactor} is tight.

\begin{example} \label{example}
Assume $n$ is a power of $2$. 
For any $z\in\{0, 1\}^{\log_2{n}}$, $e(z)\in\{0, 1\}^n$ is 
defined as follows:
$e(z)_i = i \cdot z$ (parity of bitwise product).
Consider $f(x):=z$ if $x=e(z)$, otherwise $f$ is undefined.
Then $E(f) = \log_2{n}$, while $Q(f) = 1$.
Let $H$ be the Hadamard transformation
on the $\log_2{n}$ index bits in $Q$, and $M$ acts on the 
last bit in $Q$ such that
$M |0\rangle =
\frac{1}{\sqrt{2}}(|0\rangle - |1\rangle)$. 
It is easy to verify that for any $x = e(z)$,
$$M^{-1} H O H M |x\rangle_P |\overrightarrow{0}\rangle_{Q} = 
|x\rangle_P |z, 0\rangle_Q.$$
\end{example}

\section{Proof of the main theorem}
For $0\le t \le T$, let $\phi_t(x) \in l_2(Q\cup R)$ be
the state such that
$$U_t O U_{t-1} \cdots U_0 |x\rangle_P |\overrightarrow{0}\rangle_{QR} 
= |x\rangle_P \otimes \phi_t(x).$$ 
Let $\Psi$ be any 
orthonormal basis for $l_2(Q\cup R)$.
Our proof will finally make use of the following
fact observed in \cite{BealsBCMW98}:
\begin{fact}
\label{reduction}
$$\phi_t(x) = \sum_{\psi \in \Psi} p_\psi(x) |\psi\rangle_{QR},$$
for some set of multi-linear polynomials $p_\psi(x)$, each of which
is of degree no more than $t$.
\end{fact}

Therefore, proving lower bounds in quantum complexity can be
reduced to proving  lower bounds on the degree of approximating polynomials.
We shall first prove some lemmas on the latter.

For any $g : \{ 0, 1 \}^n \rightarrow \mathbf{R}$, 
define the average sensitivity of $g$, 
$$\bar{s}_g = E_{x, i}\left[ (g(x) - g(x+e_i))^2 \right]\ \mathrm{and,}$$ 

$$p_g = E_x\left[ g(x) \right].$$ All randomness is uniform.
When $g$ is a Boolean function, $\bar{s}_g$ is just
the probability that a random edge in the Boolean cube connects two vertices of 
different function values, and $p_g$ is the probability for a 
random input to have function value $1$. 

Now let $g$ be a Boolean function, and $\tilde{g}: \{0, 1\}^n \rightarrow [0, 1]$ 
approximate $g$, i.e., $|\tilde{g}(x) - g(x)| \le 1/3$ for all $x \in \{0, 1\}^n$.
The following theorem says
that a larger $\bar{s}_g$ or a smaller $p_{\tilde{g}}$ will force $\tilde{g}$
to have high degree.

\begin{lemma}\label{smallaverage}  $deg(\tilde{g}) \ge n\bar{s}_{\tilde{g}}/(4p_{\tilde{g}})
\ge n\bar{s}_g/(36p_{\tilde{g}})$.
\end{lemma}

\begin{proof} Let $d = deg(\tilde{g})$, then the Fourier representation of $\tilde{g}$ is
$$\tilde{g}(x) = \sum_{r\in\{0, 1\}^n, |r| \le d} \hat{\tilde{g}}_{r} (-1)^{x \cdot r},$$  
where $\hat{\tilde{g}}_r = \mathrm{E}_x [\tilde{g}(x) (-1)^{x \cdot r}]$.
By simple calculation,
$$\bar{s}_{\tilde{g}} = \sum_{r, |r| \le d} \hat{\tilde{g}}_r^2 \frac{4|s|}{n} \le
(\sum_{r, |r| \le d} \hat{\tilde{g}}_r^2) \frac{4d}{n} =
\mathrm{E}_x\left[ \tilde{g}^2(x) \right]4d/{n} \le 4d{p_{\tilde{g}}}/{n}.$$ 
Since $\tilde{g}$ approximates $g$, $\bar{s}_{\tilde{g}} \ge \frac{1}{9}\bar{s}_g$.
\end{proof}

The following lemma about Boolean functions will be needed 
immediately:

\begin{lemma}\label{edgebound} Let $k$ be the cardinality of 
$X \subseteq \{ 0, 1\}^n$, $t_X$  the
number of edges in the Boolean cube that connect two 
vertices in $X$. Then $t_X \le k\log_2{k}/2$.
\end{lemma}

\begin{proof} By induction. 
It is true for $k = 1, 2$. Assume the statement is 
true for all natural numbers smaller than $k$, and let's examine the case
$k \ge 3$. Pick a coordinate $i$ such that both the subcubes of $x_i = 1$ and $x_i = 0$ have 
nonempty subsets $A$ and $B$ of $X$. 
Then $t_X \le t_A + t_B + \min\{ |A|, |B| \}$. 
We can assume without loss of generality
that $1 \le |A| = a \le k/2$. Then by simple calculation,
$$t_X \le \frac{1}{2}a\log_2{a} + \frac{1}{2}( k - a )\log_2{( k - a )} + a \le k\log_2{k}/2.$$
\end{proof}

Let $H(\cdot)$ be the entropy function, i.e., for $\eta\in[0, 1]$,
$H(\eta) := \eta\log_2{\frac{1}{\eta}} + (1 - \eta)\log_2{\frac{1}{1 - \eta}}$.
The following lemma says that if the number of true assignments is close to
the number of false assignments,
then the Boolean function should have high average sensitivity:

\begin{lemma}\label{highsensitivity} For any Boolean function $g$,
$\bar{s}_g \ge H(p_g)/n$.
\end{lemma}

\begin{proof} Let $k = 2^n p_g$ be the number of true assignments. 
By Lemma \ref{edgebound}, in the Boolean cube, the number of edges that
connect two true assignments is less than $k\log_2{k}/2$, and the number of edges
that connect two false assignments is less than $(2^n - k) \log_2{(2^n - k)}/2$.
Therefore, 
$$\bar{s}_g = \Pr_{x, i} \left[ g(x) \ne g(x+e_i) \right]
\ge \left( n2^n - k\log_2{k} - (n-k)\log_2{(n-k)} \right)/n2^n = H(p_g)/n.$$
\end{proof}

We are now ready to prove our main theorem:

\begin{proof}[\textbf{Main Theorem}]
For each $y\in B$, 
let $f_y$ be the characteristic function of $f^{-1}(y)$, i.e., 
\begin{displaymath}
f_y(x) = \left\{
	\begin{array}{ll}
           1&\textrm{if $f(x) = y$,}\\
	   0&\textrm{otherwise.}
	\end{array}
	\right.
\end{displaymath}
Let $\tilde{f}_y(x)$
be the probability that $y$ is observed 
as the output when the input is $x$. Then by Lemma~\ref{reduction},
$\tilde{f}_y$ is a nonnegative 
polynomial of degree no more than $2Q(f)$, and
$\tilde{f}_y$ approximates $f_y$.
Furthermore, for any $x$, $\sum_y \tilde{f}_y(x) \le 1$.

For simplicity of notation, we shall use $p_y$ in place 
for $p_{f_y}$, $\tilde{p}_y$ for $p_{\tilde{f}_y}$, and
$\bar{s}_y$ for $\bar{s}_{f_y}$. 
Note that $$E(f) = \sum_y p_y\log_2{\frac{1}{p_y}},$$
and,
$$\sum_y\tilde{p}_y \le 1.$$ 
Let $d = \max_y deg(\tilde{f}_y)$. We want to get
a lower bound for $d$.

By Lemma \ref{smallaverage} 
$$\frac{n}{36}\bar{s}_y \le d_y\tilde{p}_y \le d\tilde{p}_y.$$ 
Summing over all $i$, and by Lemma~\ref{highsensitivity}, 
we get 
$$d \ge \frac{n}{36}\sum_y\bar{s}_y \ge \frac{1}{36} \sum_y{H(p_y)}
\ge \frac{1}{36} E(f).$$
\end{proof}

\section{Acknowledgment}
The author would like to thank Andy~Yao for discussion, 
Jason~Perry and Lane~Hemaspaandra for going through the
paper and giving useful comments.

\end{document}